\documentclass{PoS}

\newcommand{\Slash}[1]{{\ooalign{\hfil/\hfil\crcr$#1$}}}

\title{
  Dirac-mode expansion analysis for Polyakov loop
}  

\ShortTitle{instantaneous 
interquark potential in generalized Landau gauge}

\author{\speaker{Takumi Iritani}, Shinya Gongyo, and Hideo Suganuma \\
        Department of Physics, Kyoto University, Kitashirakawaoiwake, Sakyo,
        Kyoto 606-8502, Japan\\
        E-mail: \email{iritani@ruby.scphys.kyoto-u.ac.jp}
}

\abstract{
  To clarify the relation between chiral symmetry breaking
  and color confinement, we investigate the Polyakov loop in terms of the 
  Dirac eigenmodes in SU(3) lattice QCD.
  We analyze the low-lying (IR) and UV Dirac-mode contribution 
  to the Polyakov loop, respectively, using the Dirac-mode expansion method.
  In the confined phase, the Polyakov loop $\langle L_P \rangle$ 
  remains almost zero and $Z_3$ center symmetry is thus unbroken, 
  even after removing low-lying Dirac-modes, 
  which are responsible to chiral symmetry breaking.
  In the confined phase, the Polyakov loop $\langle L_P \rangle$ also 
  remains almost zero by UV Dirac-modes cut. 
  In addition to the confined phase, we analyze the Polyakov loop 
  in the deconfined phase and its temperature dependence. 
  The behavior of the Polyakov loop $\langle L_P \rangle$ is found to be 
  almost unchanged by the cut of low-lying or UV Dirac-modes
  in both confined and deconfined phases.
}
         
\FullConference{The 30 International Symposium on Lattice Field Theory - Lattice 2012,\\
		June 24-29, 2012\\
        Cairns, Australia}

\begin{document}

\section{Introduction}
  Nowadays, quantum chromodynamics (QCD) 
  is established as the fundamental theory of the strong interaction.
  However, the non-perturbative properties of QCD
  are not fully understood, especially, 
  on color confinement and chiral symmetry breaking.
  It is interesting issue to 
  investigate the correspondence between these non-perturbative phenomena
  \cite{Suganuma:1995,Miyamura:1995,Gattringer:2006,
  Bruckmann:2007,Bilgici:2008,Synatschke:2008,Suganuma:2011,Gongyo:2012}.
  The lattice QCD calculations show 
  the simultaneous chiral and deconfined phase transitions at finite temperature \cite{Karsch:2002}, which 
  suggests a close relation between confinement and chiral 
  symmetry breaking.

  As for chiral symmetry breaking,
  the chiral condensate $\langle \bar{q} q \rangle$
  is directly connected to the Dirac operator in QCD.
  The chiral condensate is proportional to the Dirac zero-mode density as
  \begin{equation}
    \langle \bar{q} q \rangle
    = - \lim_{m \rightarrow 0}
    \lim_{V\rightarrow \infty} \pi \langle \rho(0) \rangle,
    \label{}
  \end{equation}
  with the Dirac spectral density $\rho(\lambda)$,
  which is known as the Banks-Casher relation \cite{BanksCasher}.
  The Dirac zero-modes are also related to the topological charge 
  via the Atiyah-Singer index theorem \cite{AtiyahSinger}.

  Therefore, it is interesting to investigate
  color confinement in terms of the Dirac-operator properties.
  Using Gattringer's formula \cite{Gattringer:2006},
  the Polyakov loop was analyzed by the Dirac spectrum sum 
  with twisted boundary condition 
  on lattice \cite{Bruckmann:2007,Bilgici:2008,Synatschke:2008}.
  In our previous studies \cite{Suganuma:2011,Gongyo:2012},
  we developed the Dirac-mode expansion method
  for the link-variable, and analyzed the role of the Dirac mode to 
  the Wilson loop and the interquark potential.
  As for the hadron spectra, it is reported that
  the hadrons still exist as the bound state even without chiral symmetry breaking 
  by removing low-lying Dirac-modes \cite{Lang:2011,Glozman:2012}.

  In this paper, 
  based on the Dirac-mode expansion method \cite{Suganuma:2011,Gongyo:2012},
  we investigate the role of the Dirac mode to
  the Polyakov loop in both confined and deconfined phases 
  at finite temperature in SU(3) lattice QCD.
  In Sec.2, we briefly review the formalism of the Dirac-mode
  expansion method in lattice QCD.
  In Sec.3, we show the lattice QCD results.
  Section 4 is devoted for the summary.

\section{Formalism}
  In this section, we briefly review the Dirac-mode expansion method 
  in lattice QCD \cite{Suganuma:2011,Gongyo:2012}, 
  and formulation of the Dirac-mode projected Polyakov loop.

  \subsection{Dirac-mode expansion in lattice QCD}
  In lattice QCD, the Dirac operator $\Slash{D} = \gamma_\mu D_\mu$ is expressed as
  \begin{equation}
    \Slash{D}_{x,y} \equiv \frac{1}{2a}
    \sum_{\mu = 1}^4
    \gamma_\mu \left[ U_\mu(x) \delta_{x+\hat{\mu},y}
    - U_{-\mu}(x) \delta_{x-\hat{\mu},y}\right],
  \end{equation}
  using the link-variable $U_\mu(x) \in \mathrm{SU}(N_c)$
  and the lattice spacing $a$.
  Here, we use the convenient notation of $U_{-\mu}(x) \equiv U_\mu^\dagger(x-\hat{\mu})$, 
  and $\hat{\mu}$ denotes for the unit vector on lattice in $\mu$-direction.
  In this paper, we adopt hermitian $\gamma$-matrices, 
  i.e., $\gamma_\mu^\dagger = \gamma_\mu$, in the Euclidean space-time.
  Thus, the Dirac operator $\Slash{D}$ is anti-hermitian,
  and the Dirac eigenvalues are pure imaginary number.
  We introduce the normalized Dirac eigenstate $|n \rangle$ which satisfies
  \begin{equation}
    \Slash{D}|n\rangle = i\lambda_n | n \rangle,
  \end{equation}
  with the eigenvalue $\lambda_n \in \mathbf{R}$.
  The Dirac eigenfunction $\psi_n(x)$ defined by
  \begin{equation}
    \psi_n(x) \equiv \langle x | n \rangle
    \label{}
  \end{equation}
  satisfies $\Slash{D}\psi_n = i \lambda_n \psi_n$.

  We consider the operator formalism in lattice QCD \cite{Suganuma:2011,Gongyo:2012}.
  The link-variable operator $\hat{U}_\mu$ is defined 
  by the matrix element
  \begin{equation}
    \langle x | \hat{U}_\mu | y \rangle = U_{\mu} (x) \delta_{x+\hat{\mu},y}.
    \label{}
  \end{equation}
  The Dirac-mode matrix element $\langle n | \hat{U}_\mu | m\rangle$
  is expressed as
  \begin{eqnarray}
    \langle n | \hat{U}_\mu | m \rangle
    &=& \sum_x \langle n | x \rangle \langle x |
    \hat{U}_\mu | x + \hat{\mu} \rangle
    \langle x + \hat{\mu} | m \rangle \nonumber \\
    &=& \sum_x \psi_n^\dagger(x) U_\mu(x) \psi_m(x+\hat{\mu}),
    \label{}
  \end{eqnarray}
  using the link-variable $U_\mu(x)$ and the Dirac eigenfunction $\psi_n(x)$.

  Using the completeness relation $\sum_n |n \rangle \langle n|=1$, 
  any operator $\hat O$ can be expanded 
  in terms of the Dirac-mode basis $|n \rangle$ as
  \begin{eqnarray}
    \hat O=\sum_n \sum_m |n \rangle \langle n|\hat O|m \rangle \langle m|,
    \label{eq:dirac-mode-expansion}
  \end{eqnarray}
  which is the mathematical basis of 
  the Dirac-mode expansion \cite{Suganuma:2011,Gongyo:2012}.

  Based on the expansion in Eq.(\ref{eq:dirac-mode-expansion}), 
  we introduce the Dirac-mode projection operator $\hat{P}$ as
  \begin{equation}
    \hat{P} \equiv \sum_{n \in \mathcal{A}} | n \rangle \langle n |,
    \label{}
  \end{equation}
  with the Dirac eigenstate $|n\rangle$,
  and arbitrary set of eigenmode subspace $\mathcal{A}$.
  For example, the IR and the UV Dirac-mode cut are given by
  \begin{equation}
    \hat{P}_{\rm IR} \equiv \sum_{|\lambda_n| \geq \Lambda_{\rm IR}} | n \rangle \langle n |, 
    \qquad
    \hat{P}_{\rm UV} \equiv \sum_{|\lambda_n| \leq \Lambda_{\rm UV}} | n \rangle \langle n |,
  \end{equation}
  respectively, with the IR/UV cutoff parameter, 
  $\Lambda_{\rm IR}$ and $\Lambda_{\rm UV}$.
  We define the Dirac-mode projected link-variable operator as
  \begin{equation}
    \hat{U}_\mu^P \equiv \hat{P} \hat{U}_\mu \hat{P} 
    = \sum_{n \in \mathcal{A}} \sum_{m \in \mathcal{A}} 
    | n \rangle \langle n | \hat{U}_\mu | m \rangle \langle m |,
    \label{}
  \end{equation}
  with the projection operator $\hat{P}$.
  Using the projected link-variable $\hat{U}_\mu^P$,
  we can analyze the individual contribution
  of each Dirac eigenmode to the various quantities, 
  such as the Wilson loop \cite{Suganuma:2011,Gongyo:2012}.

  \subsection{Polyakov loop operator and Dirac-mode projection}
  Next, we formulate the Dirac-mode projection of the Polyakov loop.
  Hereafter, we consider the periodic SU(3) lattice of 
  the space-time volume $V = L^3 \times N_t$ with lattice spacing $a$.
  In lattice QCD operator formalism,
  the Polyakov-loop operator is defined by
  \begin{equation}
    \hat{L}_P \equiv \frac{1}{3V} \prod_{i=1}^{N_t} \hat{U}_4 
             =\frac{1}{3V} \hat{U}_4^{N_t}
    \label{}
  \end{equation}
  with the temporal link-variable operator $\hat{U}_4$.
  Taking the functional trace ``Tr'',
  we obtain the standard Polyakov loop $\langle L_P \rangle$ as
  \begin{eqnarray}
    \mathrm{Tr} \ \hat{L}_P 
    &=& \frac{1}{3V} \mathrm{Tr} \ 
    \large\{ \prod_{i=1}^{N_t} \hat{U}_4 \large\} = \frac{1}{3V} \mathrm{tr}
    \sum_{\vec{x},t} \langle \vec{x},t | \prod_{i=1}^{N_t} \hat{U}_4 | \vec{x},t \rangle \nonumber  \\
    &=& \frac{1}{3V} \mathrm{tr} \sum_{\vec{x},t} \langle \vec{x},t | \hat{U}_4 | \vec{x},t + a \rangle 
      \langle \vec{x},t+a | \hat{U}_4 | \vec{x},t+2a \rangle 
      \cdots \langle \vec{x},t + (N_t-1)a | \hat{U}_4 | \vec{x},t \rangle \nonumber \\
    &=& \frac{1}{3V} \mathrm{tr}
      \sum_{\vec{x},t} U_4(\vec{x},t) U_4(\vec{x},t+a) \cdots U_4(\vec{x},t+(N_t-1)a)
      = \langle L_P \rangle,
  \end{eqnarray}
  where ``tr'' denotes the trace over SU(3) color index.

  Using the projection operator $\hat P$,
  we define the Dirac-mode projected Polyakov loop 
  $\langle L_{P}^{\rm proj.} \rangle$ as
  \begin{eqnarray}
    \langle L_{P}^{\rm proj.} \rangle
    &\equiv& \frac{1}{3V} \mathrm{Tr} \ \large\{ \prod_{i=1}^{N_t} \hat{U}_4^P \large\} = 
    \frac{1}{3V} \mathrm{Tr} \left\{
    \hat{P}\hat{U}_4 \hat{P} \hat{U}_4 \hat{P} \cdots \hat{P} \hat{U}_4 \hat{P} \right\} \nonumber \\
    &=& \frac{1}{3V}\mathrm{tr} \sum_{n_1,n_2,\dots, n_{N_t} \in \mathcal{A}}
      \langle n_1 | \hat{U}_4 | n_2 \rangle \langle n_2 | \hat{U}_4 | n_3 \rangle
      \cdots \langle n_{N_t} | \hat{U}_4 | n_1 \rangle.
    \label{eq:Lproj}
  \end{eqnarray}
  In particular, we consider the IR/UV Dirac-mode projected-Polyakov loop as
  \begin{eqnarray}
    \langle L_P \rangle_{\rm IR}
    &\equiv& \frac{1}{3V} \mathrm{tr}
    \sum_{|\lambda_{n_i}| \ge \Lambda_{\rm IR}}
    \langle n_1 | \hat{U}_4 | n_2 \rangle
    \cdots \langle n_{N_t} | \hat{U}_4 | n_1 \rangle, \\
    \langle L_P \rangle_{\rm UV}
    &\equiv& \frac{1}{3V} \mathrm{tr}
    \sum_{|\lambda_{n_i}| \le \Lambda_{\rm UV}}
    \langle n_1 | \hat{U}_4 | n_2 \rangle
    \cdots \langle n_{N_t} | \hat{U}_4 | n_1 \rangle,
    \label{}
  \end{eqnarray}
  with the IR/UV eigenvalue cutoff, $\Lambda_{\rm IR}$ and $\Lambda_{\rm UV}$.

\section{Lattice QCD calculation}
  In this section, we study 
  the Polyakov loop in terms of the Dirac-mode 
  in SU(3) lattice QCD at the quenched level.
  We use the LAPACK package for the full diagonalization of the Dirac operator \cite{LAPACK}.
  We use the Kogut-Susskind (KS) formalism for reduction of the computational costs
  \cite{Suganuma:2011,Gongyo:2012}.

  \subsection{The confined phase}
\begin{figure}
  \centering
  \includegraphics[width=0.325\textwidth,clip]{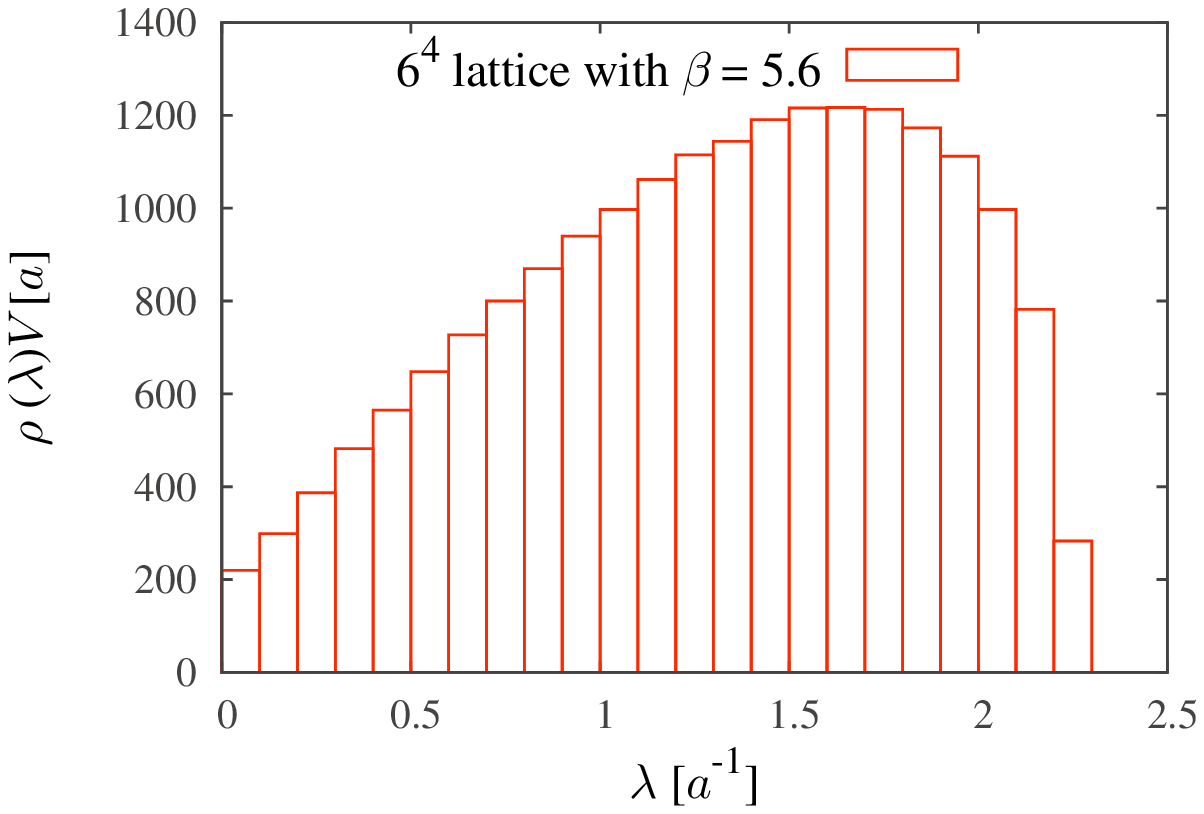}
  \includegraphics[width=0.325\textwidth,clip]{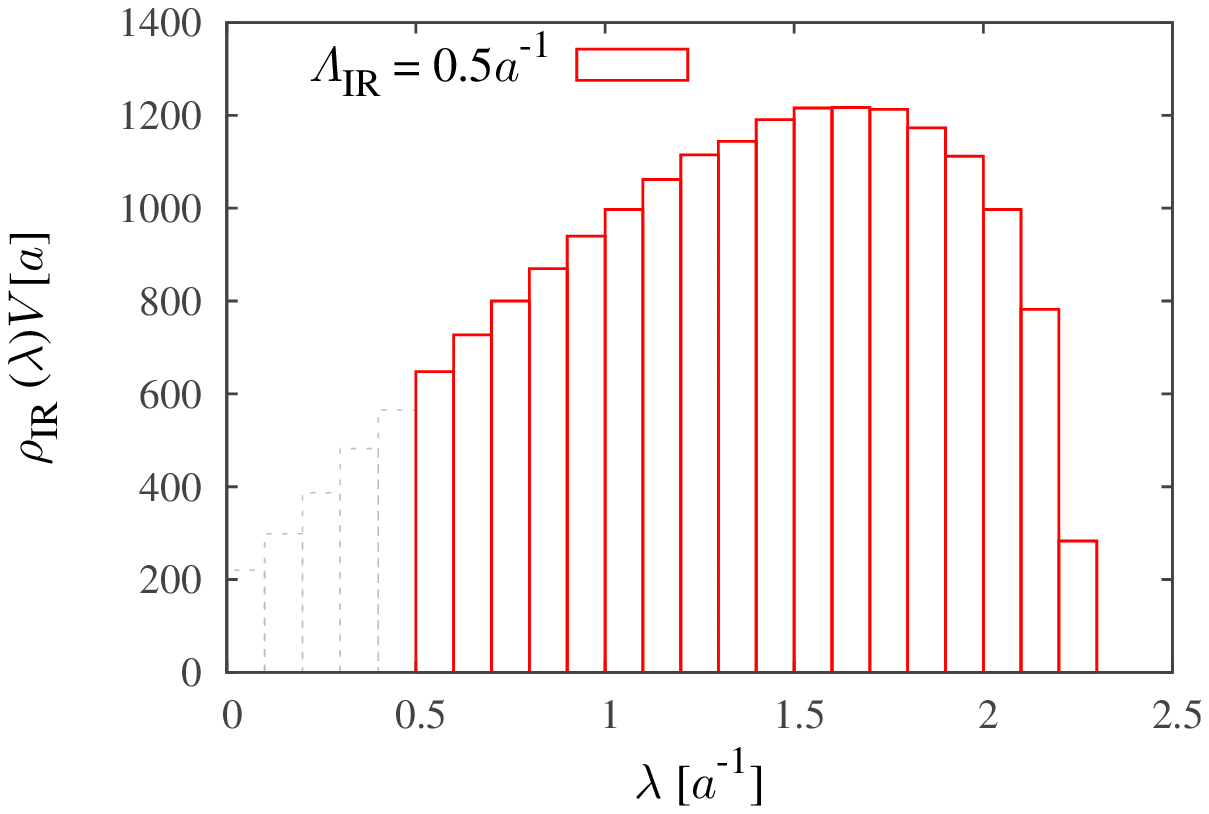}
  \includegraphics[width=0.325\textwidth,clip]{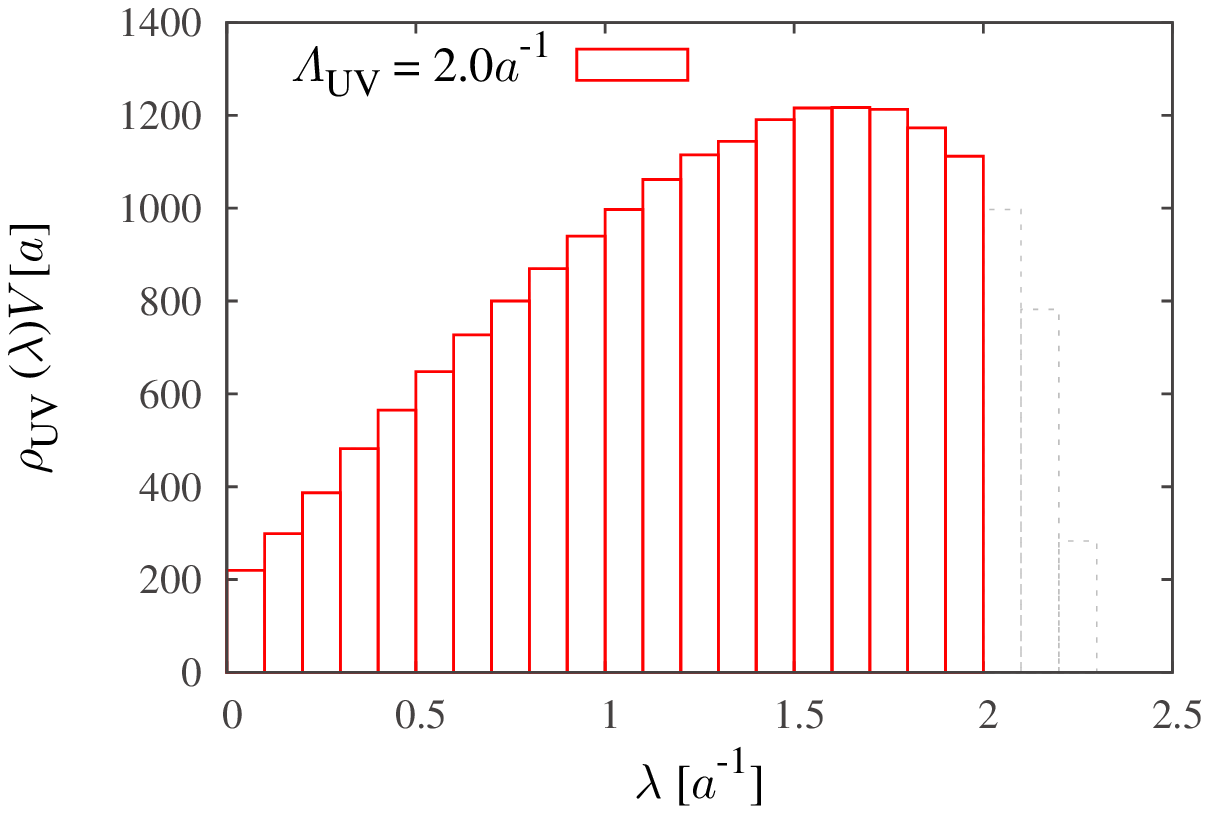}
  \caption{\label{fig:DiracSpectrumCut} 
    The Dirac spectral density in SU(3) lattice QCD 
    on $6^4$ at $\beta$=5.6, i.e., $a\simeq$0.25fm.
  (a) The original spectral density $\rho(\lambda)$. 
  (b) $\rho_{\rm IR}(\lambda)$ for IR-cut with $\Lambda_{\rm IR} = 0.5a^{-1}$.
  (c) $\rho_{\rm UV}(\lambda)$ for UV-cut with $\Lambda_{\rm UV} = 2.0a^{-1}$.
  }
\end{figure}

  First, we analyze the Polyakov loop properties in the confined phase.
  Here, we use $6^4$ lattice with $\beta = 5.6$, which
  corresponds to lattice spacing $a \simeq 0.25$fm \cite{Suganuma:2011,Gongyo:2012}.
  The total number of KS Dirac-modes is $L^3 \times N_t \times 3 = 3888$.
  Figure \ref{fig:DiracSpectrumCut} shows
  the lattice QCD result for the Dirac spectral density $\rho(\lambda)$ 
  and IR/UV-cut Dirac spectral density, 
  \begin{equation}
    \rho_{\rm IR}(\lambda) \equiv \rho(\lambda)\theta(|\lambda|-
    \Lambda_{\rm IR}), 
    \quad
    \rho_{\rm UV}(\lambda) \equiv \rho(\lambda)
    \theta( \Lambda_{\rm UV} - | \lambda | ), 
  \end{equation}
  with $\Lambda_{\rm IR} = 0.5a^{-1}$ and $\Lambda_{\rm UV} = 2.0a^{-1}$.
  Both mode-cuts correspond to removing about 400 modes from full eigenmodes.

\begin{figure}
  \centering
  \includegraphics[width=0.27\textwidth,clip]{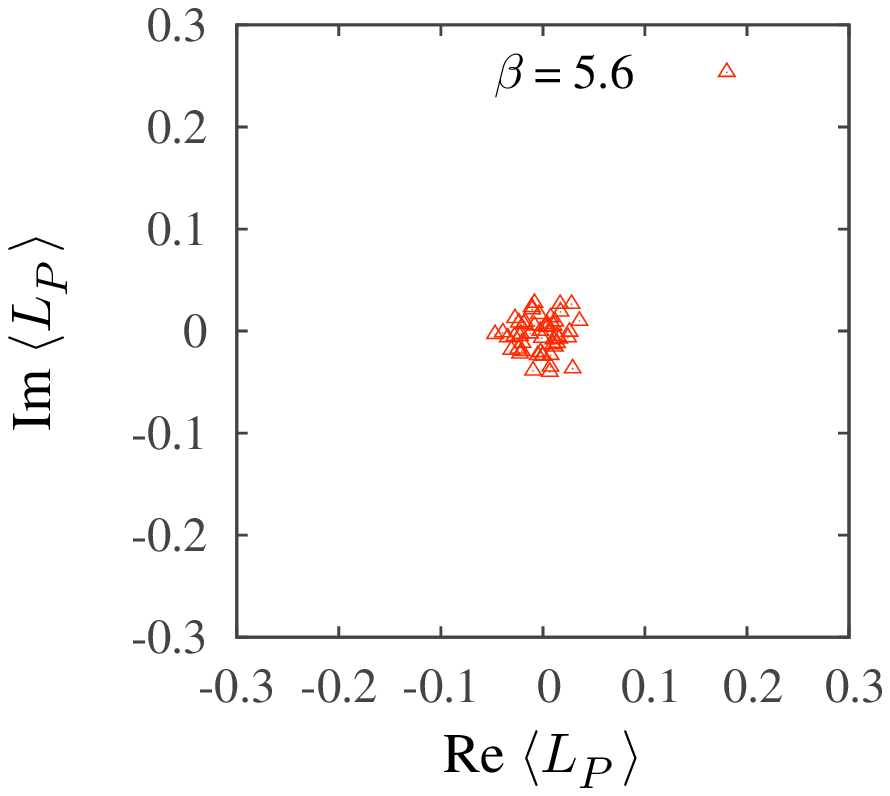}
  \includegraphics[width=0.27\textwidth,clip]{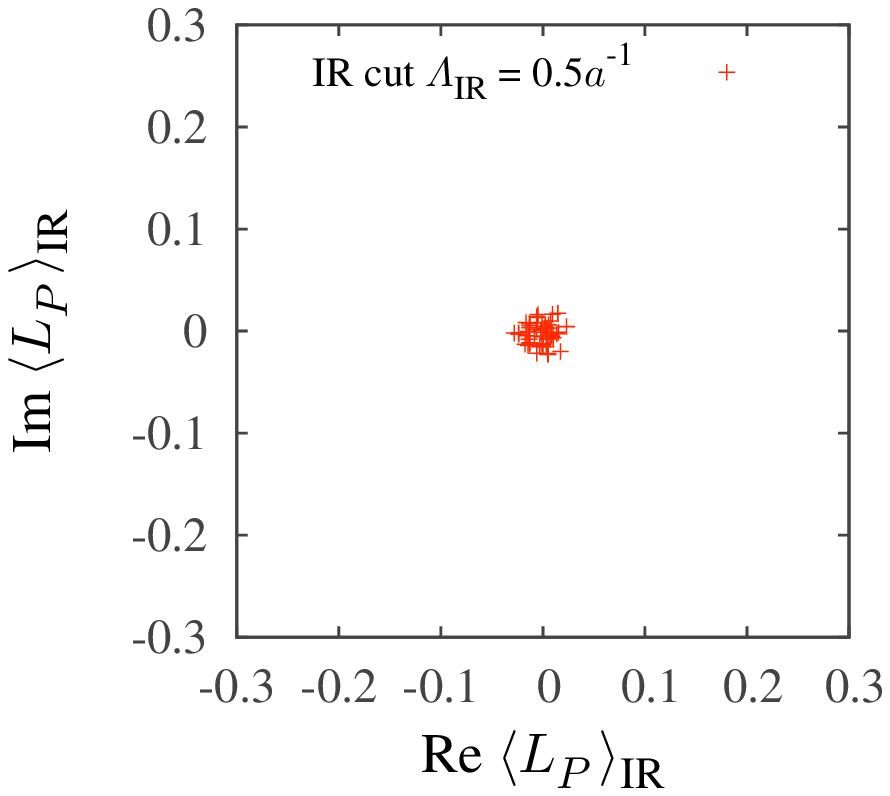}
  \includegraphics[width=0.27\textwidth,clip]{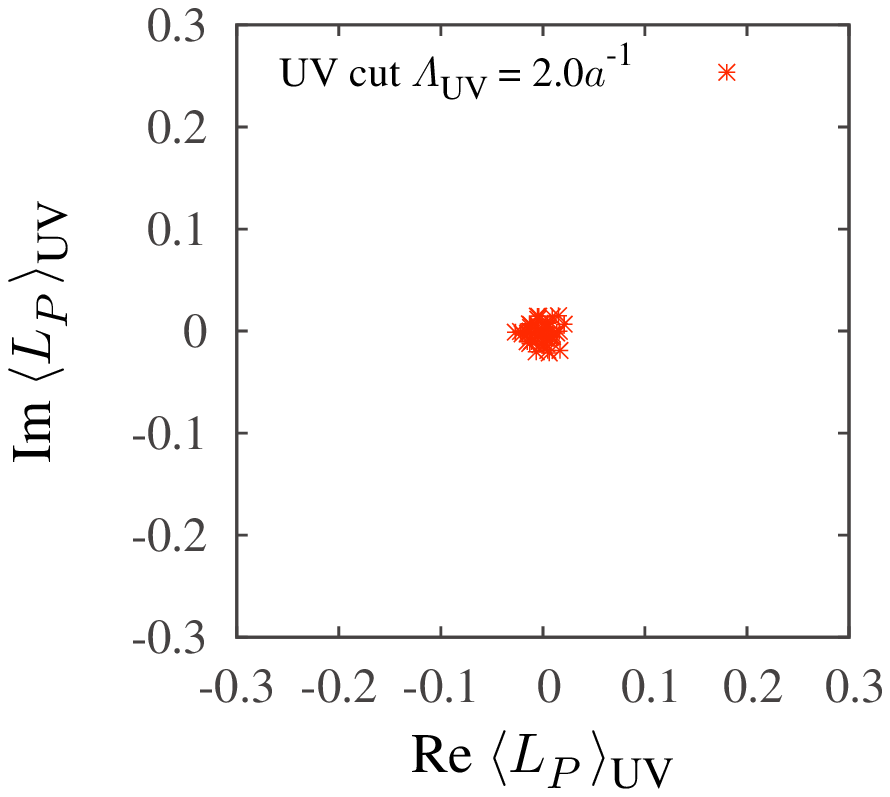}
\vspace{-0.2cm}
  \caption{\label{fig:PolyakovScatterConfined} 
    The scatter plot of the Polyakov loop in the confined phase 
    in SU(3) lattice QCD on $6^4$ at $\beta$=5.6, i.e.,
    $a\simeq$0.25fm and $T \equiv 1/(N_ta) \simeq$ 0.13GeV.
    (a) The original Polyakov loop $\langle L_P \rangle$.
    (b) $\langle L_P \rangle_{\rm IR}$ for low-lying Dirac-mode cut
        with $\Lambda_{\rm IR} = 0.5a^{-1}$.
    (c) $\langle L_P \rangle_{\rm UV}$ for UV Dirac-mode cut 
        with $\Lambda_{\rm UV} = 2.0a^{-1}$.
  }
\end{figure}

  We show in Figs.\ref{fig:PolyakovScatterConfined}(a)-(c) 
  the scatter plot of the original Polyakov loop $\langle L_P \rangle$, 
  $\langle L_P \rangle_{\rm IR}$ for low-lying Dirac-mode cut 
  with $\Lambda_{\rm IR} = 0.5a^{-1}$, 
  and $\langle L_P \rangle_{\rm UV}$ for UV Dirac-mode cut 
  with $\Lambda_{\rm UV} = 2.0a^{-1}$, respectively.
  As shown in Fig.\ref{fig:PolyakovScatterConfined}(a),
  the Polyakov loop satisfies $\langle L_P \rangle \simeq 0$,
  which indicates the confined phase.

  By removing low-lying Dirac-modes, 
  chiral symmetry breaking is effectively restored 
  \cite{Gongyo:2012,BanksCasher,Lang:2011,Glozman:2012}.
  Actually, this IR Dirac-mode cut of 
  $\Lambda_{\rm IR} = 0.5a^{-1} \simeq 0.4$GeV  
  corresponds to about 98\% reduction of the quark condensate
  around the physical region $m_q \simeq 5$MeV \cite{Gongyo:2012}.
  However, as shown in Fig.\ref{fig:PolyakovScatterConfined}(b), 
  the Polyakov loop $\langle L_P \rangle_{\rm IR}$ remains almost zero, 
  which means unbroken $Z_3$ center symmetry.
  This result indicates that 
  the single-quark energy is still extremely large, and the system remains 
  in the confined phase even without chiral symmetry breaking.
  In the UV Dirac-mode cut, the chiral condensate is almost unchanged, 
  and the Polyakov loop $\langle L_P \rangle_{\rm UV}$ 
  also remains almost zero, 
  as shown in Fig.\ref{fig:PolyakovScatterConfined}(c).
  These results in Figs.\ref{fig:PolyakovScatterConfined}(b) and (c)
  show that the Polyakov loop is insensitive to the IR/UV Dirac-mode cut.

\subsection{The deconfined phase at high temperature}
  Next, we study the role of the Dirac mode 
  in the deconfined phase at high temperature.
  Here, we use $6^3 \times 4$ lattice at $\beta$=6.0.
  The total number of KS Dirac eigenmodes is $L^3 \times N_t \times 3$=2592.

\begin{figure}
    \centering
    \includegraphics[width=0.27\textwidth,clip]{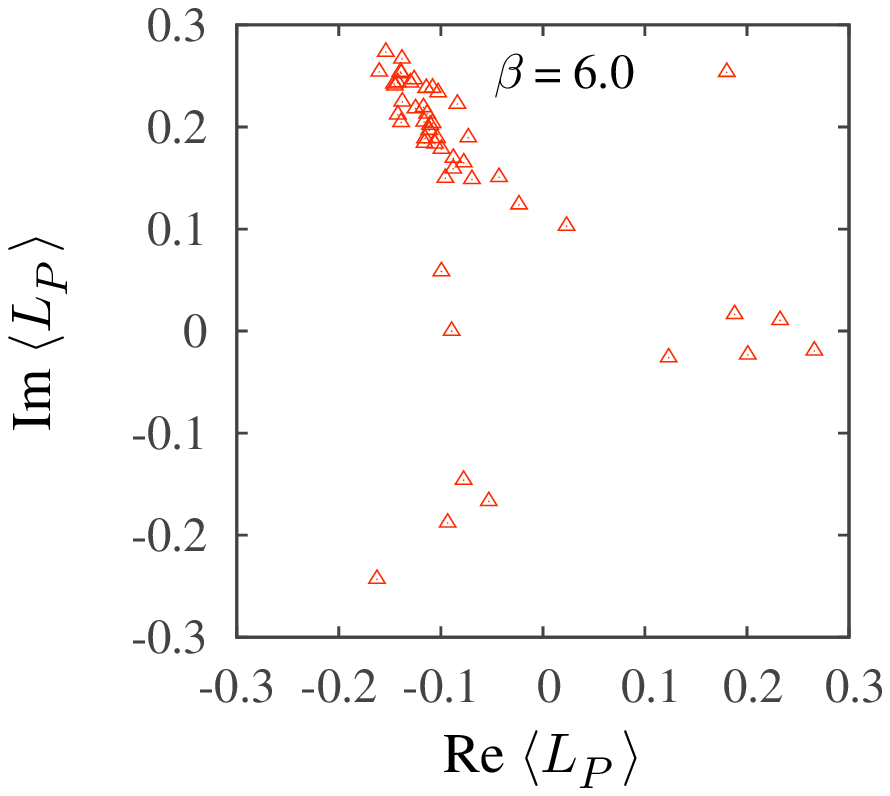}
    \includegraphics[width=0.27\textwidth,clip]{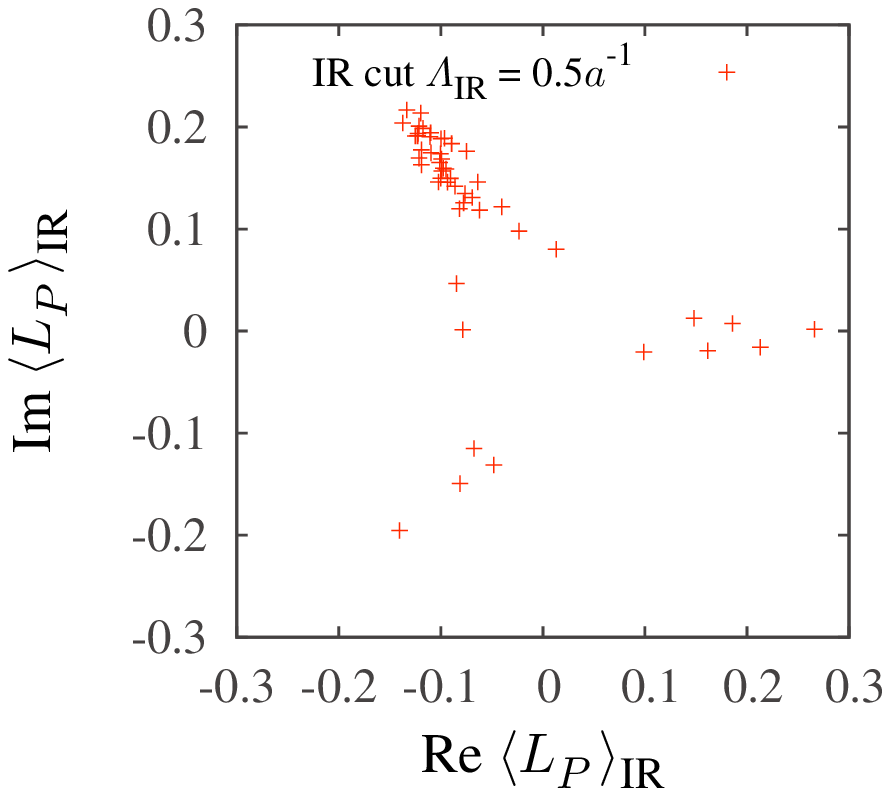}
    \includegraphics[width=0.27\textwidth,clip]{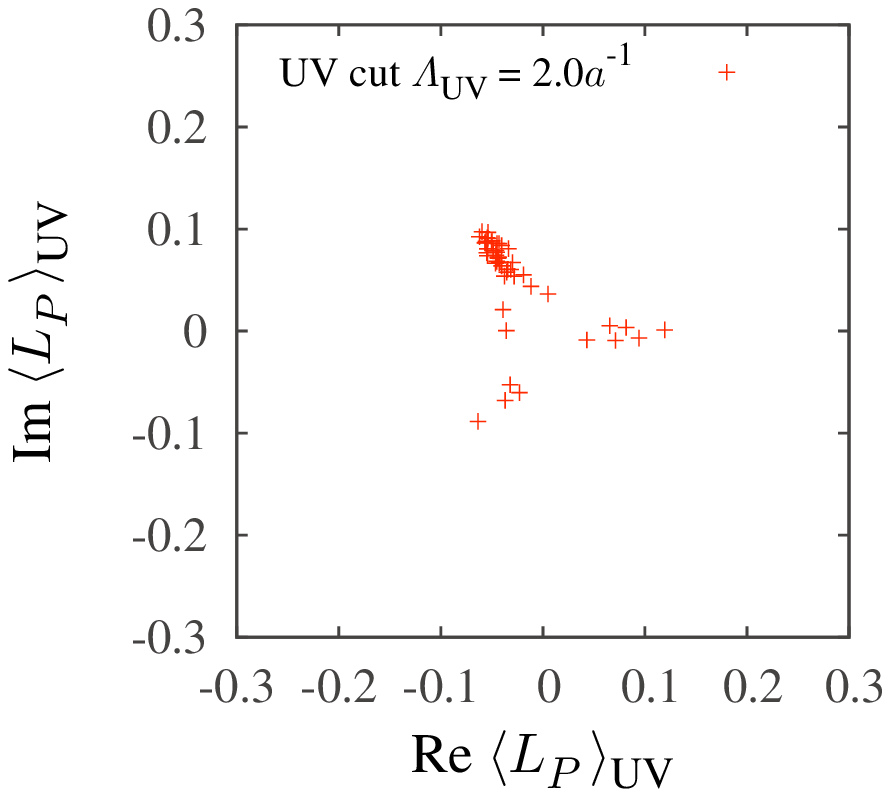}
\vspace{-0.2cm}
    \caption{\label{fig:PolyakovScatterDeconfined}
    The scatter plot of the Polyakov loop in the deconfined phase 
    in SU(3) lattice QCD on $6^3 \times 4$ at $\beta$=6.0, i.e.,
    $a\simeq$0.10fm and $T \equiv 1/(N_ta) \simeq$ 0.5GeV.
    (a) The original Polyakov loop $\langle L_P \rangle$.
    (b) $\langle L_P \rangle_{\rm IR}$ for low-lying Dirac-mode cut
        with $\Lambda_{\rm IR} = 0.5a^{-1}$.
    (c) $\langle L_P \rangle_{\rm UV}$ for UV Dirac-mode cut 
        with $\Lambda_{\rm UV} = 2.0a^{-1}$.
    }
  \end{figure}

  Figure~\ref{fig:PolyakovScatterDeconfined} shows
  the original Polyakov loop $\langle L_P \rangle$, 
  $\langle L_P \rangle_{\rm IR}$ for low-lying Dirac-mode cut
        with $\Lambda_{\rm IR} = 0.5a^{-1}$, and 
  $\langle L_P \rangle_{\rm UV}$ for UV Dirac-mode cut 
        with $\Lambda_{\rm UV} = 2.0a^{-1}$, respectively.
  These mode-cuts correspond to removing about 200 modes from full eigenmodes.
  As shown in Fig.\ref{fig:PolyakovScatterDeconfined}(a),
  the Polyakov loop has a non-zero expectation value 
  $\langle L_P \rangle \neq 0$, 
  which shows the center group $Z_3$ structure on the complex plane.
  This property indicates the deconfined phase. 

  After removing low-lying or UV Dirac-modes, 
  as shown in Figs.\ref{fig:PolyakovScatterDeconfined}(b) and (c),
  the Dirac-mode projected Polyakov loop $\langle L_P \rangle_{\rm IR/UV}$ 
  still shows the non-zero value and the $Z_3$ center structure, 
  which indicates the deconfined and $Z_3$ broken phase.
  In fact, in both cut cases of IR and UV Dirac modes, 
  no drastic change occurs on the Polyakov loop, 
  apart from a constant normalization factor. 
  The Dirac-mode seems to be insensitive also for deconfinement properties of 
  the Polyakov loop.

\subsection{$\beta$-dependence of Dirac-mode projected Polyakov loop}
  We also investigate the $\beta$-dependence of the absolute value 
  of the Dirac-mode projected Polyakov loop, 
  $|\langle L_{\rm P} \rangle_{\rm IR/UV}|$, 
  at fixed $N_t$ and $L$.
  Here, we use $6^3 \times 4$ lattice with $\beta = 5.4 \sim 6.0$.

\begin{figure}
  \begin{center}
    \includegraphics[width=0.40\textwidth,clip]{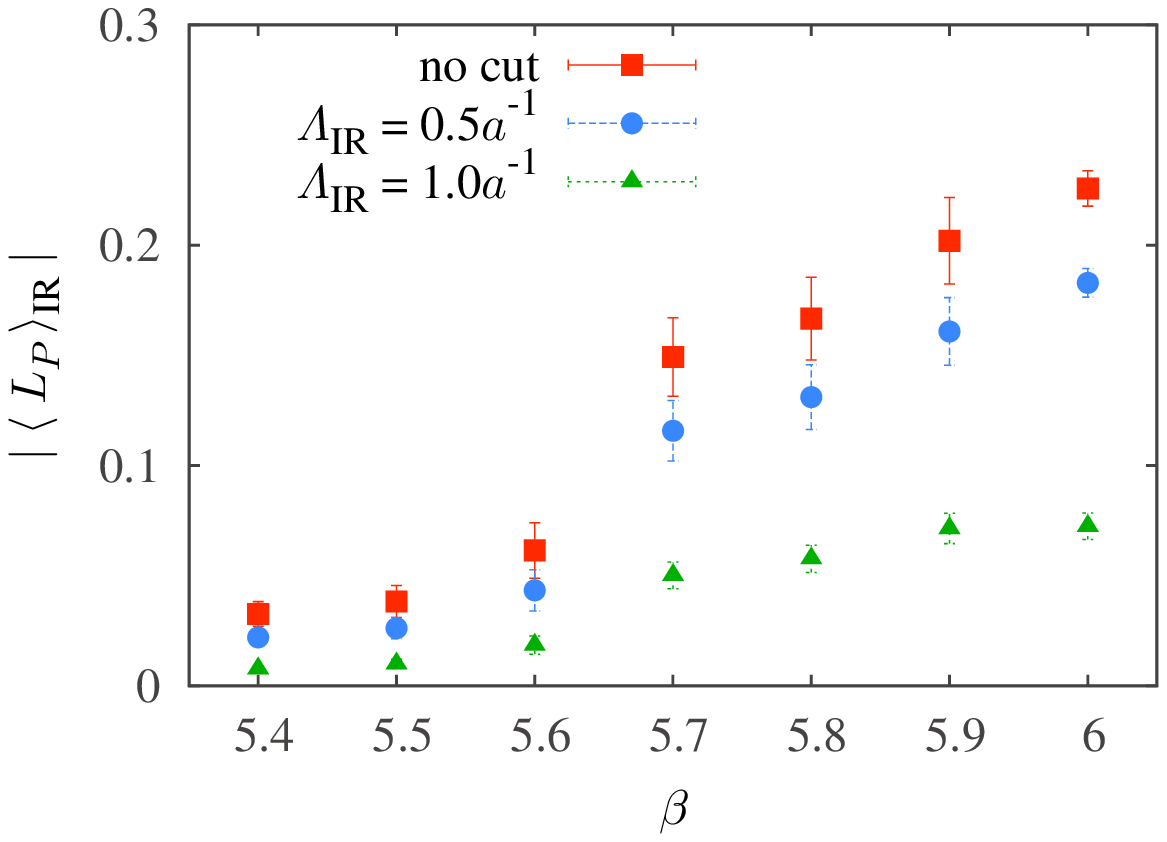}
    \includegraphics[width=0.40\textwidth,clip]{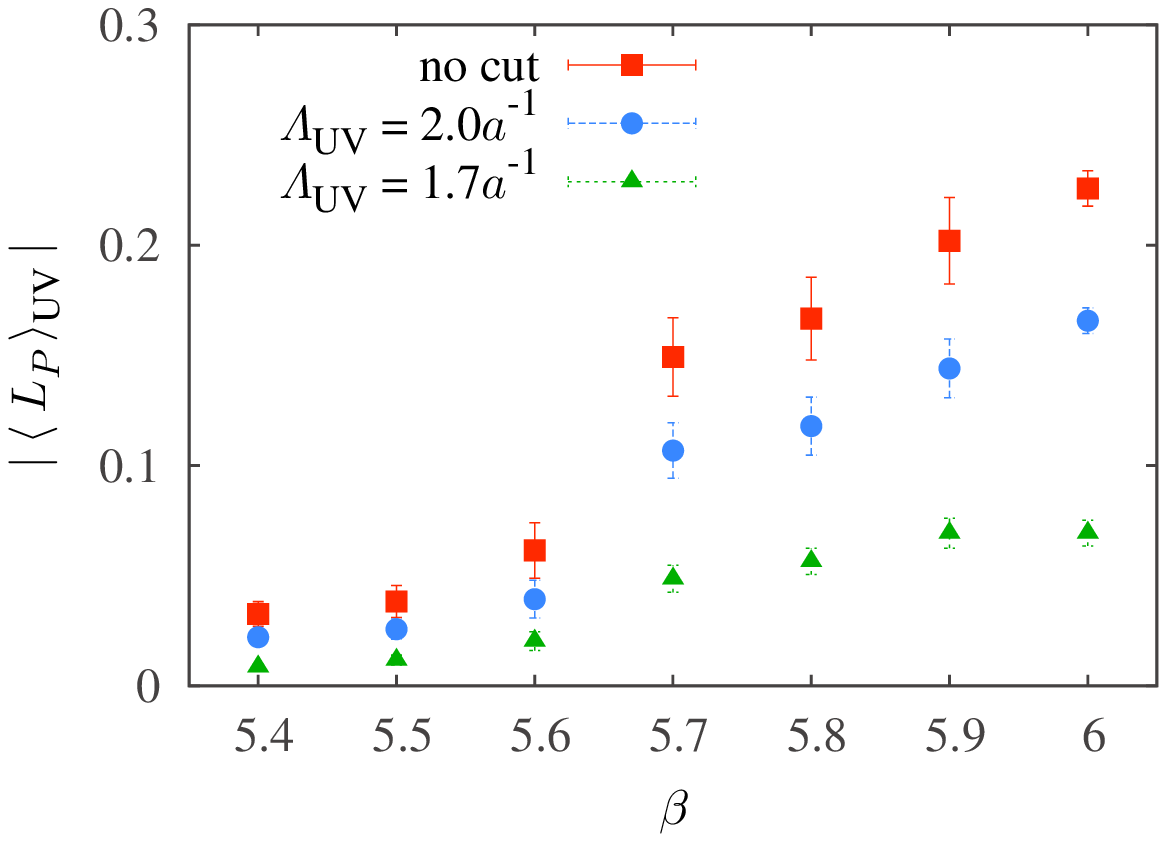}
\vspace{-0.5cm}
  \end{center}
  \caption{ \label{fig:PolyakovDiracCut}
   The $\beta$-dependence of the Polyakov loop 
   $|\langle L_{\rm P} \rangle_{\rm IR/UV}|$ in SU(3) lattice QCD on 
   $6^3 \times 4$.
   (a) $|\langle L_{\rm P} \rangle_{\rm IR}|$ for IR Dirac-mode cut 
       with $\Lambda_{\rm IR} = 0.5a^{-1}$ and $1.0a^{-1}$.
   (b) $|\langle L_{\rm P} \rangle_{\rm UV}|$ for UV Dirac-mode cut 
       with $\Lambda_{\rm UV} = 2.0a^{-1}$ and $1.7a^{-1}$.
 The original Polyakov loop $|\langle L_{\rm P} \rangle|$ without cut is added.
  }
\end{figure}

  Figure~\ref{fig:PolyakovDiracCut} (a) shows 
  $|\langle L_{\rm P} \rangle_{\rm IR}|$ 
  with $\Lambda_{\rm IR} = 0.5a^{-1}$ and $1.0a^{-1}$, and 
  Fig.\ref{fig:PolyakovDiracCut} (b) shows
  $|\langle L_{\rm P} \rangle_{\rm UV}|$
  with $\Lambda_{\rm UV} = 2.0a^{-1}$ and $1.7a^{-1}$.
  In terms of the removed number of Dirac modes, 
  $\Lambda_{\rm IR} = 0.5a^{-1}$ and $1.0a^{-1}$ 
  approximately correspond to 
  $\Lambda_{\rm UV} = 2.0a^{-1}$ and $1.7a^{-1}$, respectively.
  We have also added the original (no Dirac-mode cut) Polyakov loop 
  $|\langle L_{\rm P} \rangle|$, 
  which shows the phase transition around $\beta = 5.6 \sim 5.7$.
  Both IR and UV Dirac-mode projected Polyakov loop 
  $|\langle L_{\rm P} \rangle_{\rm IR/UV}|$ 
  show the similar $\beta$-dependence as the original Polyakov loop 
 $|\langle L_{\rm P} \rangle|$, 
  apart from a constant normalization factor.

\section{Summary and concluding remarks}
  In this paper, we have analyzed the direct relation between the Dirac eigenmodes 
  and the Polyakov loop in SU(3) lattice QCD calculation at the quenched level.
  Using the Dirac-mode expansion method,
  we have carefully removed the relevant ingredient of chiral symmetry breaking
  from the Polyakov loop.

  In the confined phase, we have found that the Polyakov loop remains almost zero 
  even without low-lying Dirac-mode.
  These low-lying modes are relevant for chiral symmetry breaking, as 
  the Banks-Casher relation indicates.
  However, the Polyakov loop does not show any drastic changes,
  which indicates the system still remains in the confined phase.
  This result is consistent with the Wilson loop analysis,
  which shows the area law and the linear interquark potential
  even after removing low-lying Dirac-modes \cite{Suganuma:2011,Gongyo:2012}.
  We have also checked the UV Dirac-mode contribution to the Polyakov loop.
  By removing UV Dirac modes, the Polyakov loop remains almost zero.
  Thus, there seem to be no specific Dirac modes 
  essential for the Polyakov loop in the confined phase.

  In addition to the confined phase, 
  we have also analyzed the Polyakov loop properties in the deconfined phase 
  at high temperature. 
  In the deconfined phase, the Polyakov loop 
  has a non-zero expectation value, 
  which distributes in $Z_3$ direction in the complex plane.
  Even by removing low-lying or UV Dirac-modes,
  the behavior of the Polyakov loop $\langle L_P \rangle_{\rm IR/UV}$ 
  seems almost unchanged, apart from a constant normalization factor.

  These lattice QCD results suggest no direct connection 
  between chiral symmetry breaking and color confinement 
  through the Dirac eigenmodes, which indicates that 
  one-to-one correspondence would not hold between them in QCD.
  If it is the case, the QCD phase diagram would exhibit more richer structure 
  by mismatch of chiral and deconfinement phase transitions.

\section*{Acknowledgements}
  The lattice QCD calculations have been done on NEC-SX8 and NEC-SX9 
  at Osaka University.
  This work is in part supported by a Grant-in-Aid for JSPS Fellows
  [No.23-752, 24-1458] and the Grant for Scientific Research 
  [(C) No.23540306, Priority Areas ``New Hadrons'' (E01:21105006)] 
  from the Ministry of Education, Culture, Science and Technology (MEXT) 
  of Japan.

\end{document}